\documentclass[aps,prx,showpacs,twocolumn,floatfix]{revtex4-2}

\usepackage{epsfig,amsmath,graphicx}
\usepackage{amssymb}
\usepackage[normalem]{ulem} 
\usepackage{braket}

\usepackage[colorlinks=true,citecolor=blue, linkcolor=blue, urlcolor=blue]{hyperref}

\usepackage[caption=false,position=top,justification=raggedright,singlelinecheck=off,labelformat=parens,subrefformat=parens,labelfont=bf]{subfig}
\usepackage[usenames,dvipsnames,svgnames,table]{xcolor}
\definecolor{NarrativeGold}{HTML}{977902}

\usepackage[usenames,dvipsnames,svgnames,table]{xcolor}
\definecolor{mypurple}{HTML}{b501b5}

\graphicspath{
{./plots/}
{./plots/final_figures/}
{./plots/rough_figures/}
}

\begin{document}


\title{Quantum Phases and Correlations Drive the Dynamics of Macroscopic Quantum Tunneling Escape in Quantum Simulators}

\author{Diego A. Alcala, Marie A. McLain, and Lincoln D. Carr}
\affiliation{$^1$Department of Physics, Colorado School of Mines, 1500 Illinois St., Golden, CO 80401, U.S.A.}

\date{\today}

\begin{abstract}
Quantum tunneling remains unexplored in many regimes of many-body quantum physics, including the effect of quantum phase transitions on tunneling dynamics. In general, the quantum phase is a statement about the ground state and has no relation to far-from-equilibrium dynamics. Although tunneling is a highly dynamical process involving many excited states, we find that the quantum phase of the Bose-Hubbard model determines phase-dependent tunneling outcomes for the quantum tunneling escape, or quasi-bound problem. Superfluid and Mott insulator correlations lead to a new quantum tunneling rate, the quantum fluctuation rate. This rate shows surprising and highly dynamical features, such as oscillatory interference between trapped and escaped atoms and a completely different macroscopic quantum tunneling behavior for superfluid and Mott insulator phases.  In the superfluid phase we find that escape dynamics are wave-like and coherent, leading to interference patterns in the density with a rapid decay process which is non-exponential. Quantum entropy production peaks when about half the atoms have escaped. In the Mott phase, despite stronger repulsive interactions, tunneling is significantly slowed by the presence of a Mott gap, creating an effective extra barrier to overcome. Only one atom can tunnel at a time, yet the decay process is nearly linear, completely defying the single-particle exponential model.  Moreover, quantum entropy peaks when only about one quarter of the atoms have escaped.  These and many other such effects go beyond the usual notions of single-particle quantum tunneling, quantum statistical effects on tunneling, and well-known semi-classical approaches from WKB to instanton theory. These results thus open up a new regime of exploration of far-from-equilibrium dynamics for quantum simulators and quantum dynamics.

\end{abstract}

\pacs{PACS}
\maketitle

\section{Introduction}
\label{sec:intro}

\begin{figure}
	\subfloat{
        \includegraphics[width=0.97\linewidth]{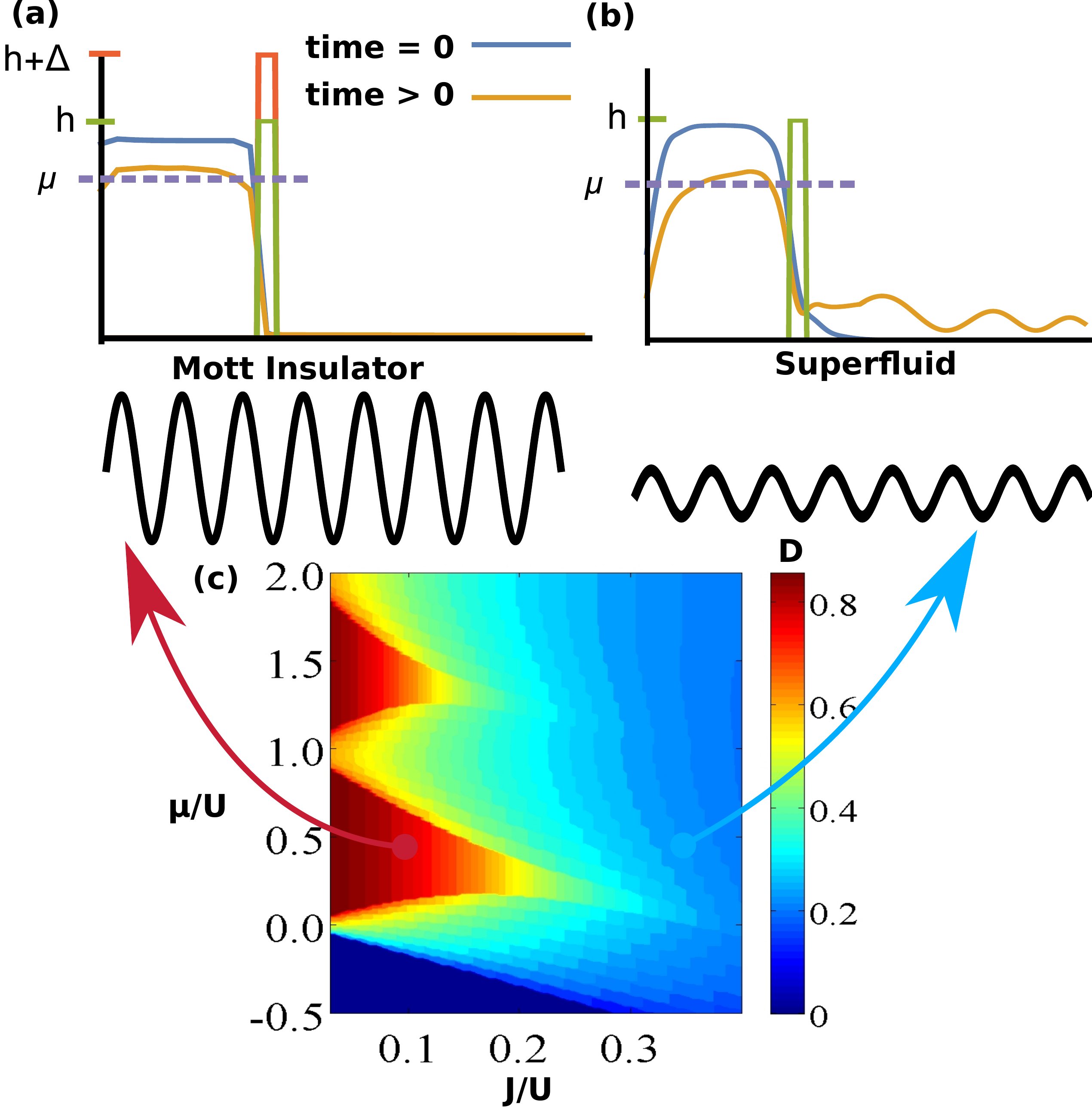}
	}
	\caption{\label{fig:Initial_State}
		\emph{Sketch of effects of the quantum phase on macroscopic quantum tunneling}
		(Upper left): The Mott insulator phase slows down tunneling due to the presence of the Mott gap over and above the effective barrier height of $h-\mu$, preventing more than one atom from escaping at a time.  (Upper right): The superfluid undergoes more rapid tunneling with characteristic wave-like interference fringes in the escape region.  (Lower panel): The mesoscopic quantum phase diagram shows the quantum depletion, or portion of the atoms outside one macroscopic semi-classical mode, as a function of chemical potential $\mu$ vs. lattice hopping energy $J$, both scaled to interaction energy $U$.  The quantum phases occupy well-defined Mott insulating lobes (left, red region) and a superfluid continuum (right, blue region), even for just $N=10$ atoms, as shown here (adapted from~\cite{carr2010mesoscopic}).
		}
\label{fig:sketch}
\end{figure}


Quantum phase transitions are the study of abrupt transitions in ground state properties of quantum matter.  In particular, a non-analyticity or singularity, typically in a correlator, occurs at zero temperature as a function of some parameter in a governing Hamiltonian.  Remarkably, the effects of this singularity emanate into the finite temperature plane as a quantum critical fan~\cite{sachdev2007quantum,carr2010understanding}.  Qualitatively speaking, this effect is like observing the effects of a gravitational singularity, or  black hole, from a distance.  A major goal of quantum simulators, or analog quantum computing devices, is to discover the phase diagrams of quantum matter, by pinpointing and characterizing quantum phase transitions.  However, many quantum simulators, such as ultracold atoms in optical lattices, are in fact much more effective at studying dynamics than statics, and thus a major application of these computing devices is uncovering the principles of far-from-equilibrium dynamics in the many-body quantum context~\cite{altman2021quantum}.  Such dynamical contexts often deviate very far from the thermal states of the quantum critical fan.  To what extent can one relate quantum phase transitions to such dynamics?  To date, the principal example of a connection between the ground state phases of quantum matter and dynamics is the Kibble-Zurek mechanism~\cite{polkovnikov2011colloquium}, in which the density of defects when ramping through a quantum phase transition is determined by the critical exponents of the quantum phase and the ramp rate.  In this Article, we discover and present a new example of the quantum phase determining dynamical outcomes far from equilibrium, namely, macroscopic quantum tunneling.

Although quantum tunneling is a well-known phenomenon in single-particle quantum physics, beginning with Gamow's 1928~\cite{Gamow1928}, and then Gurney and Condon's~\cite{Gurney1929} independent 1929 explanation of radioactive decay, both theoretical predictions and experimental demonstrations in many-body quantum physics have yet to be discovered in most regimes~\cite{zhao_macroscopic_2017}.  However, there have been a few cases of new tunneling regimes uncovered in quantum simulators.  These are dominated by either nonlinearity, a classical wave effect and very much in the semiclassical limit, or by bosonic or fermionic statistics.  Examples include observation of the tunneling to nonlinear self-trapping transition of a Bose-Einstein condensate (BEC) in a double well potential~\cite{albiez_direct_2005}; controlled tunneling escape of fermions via pairing and quantum statistics~\cite{zurn2013pairing}; and interaction-assisted escape of a BEC and emergence of a non-exponential escape rate~\cite{potnis_interaction-assisted_2017}.  A classic example is the tunneling of the mean field of an attractive BEC from a metastable to a runaway collapse process in variational space via the instanton approximation~\cite{ueda1998macroscopic}, which appears to be swamped by thermal fluctuations over the barrier and thus remains an open question of interpretation in the experimental observation of collapsing BECs (see~\cite{gerton2000direct,khaykovich2002formation,carr2004spontaneous,carr2004pulsed,cornish2006formation} for earlier work and~\cite{wilson2020dynamics} for a very recent example).  In all of these cases the interactions inherent in the quantum matter, whether bosonic or fermionic, significantly modify the tunneling dynamics.  However, quantum correlations beyond Fermi/Bose statistics and entanglement have not yet played a role.  Thus the study of macroscopic or many-body quantum tunneling has to-date mainly been constrained to those features we primarily associate with developments in physics before the era of quantum information science and tunable quantum computational devices.  By incorporating quantum phase transitions into macroscopic quantum tunneling, we take tunneling into a new regime in this Article.

A surprising fact about quantum phase transitions is that despite the Mermin-Wagner-Hohenbeg theorem demonstrating lack of a limit to a perfect non-analyticity in one dimension, nevertheless 1D systems such as quasi-1D BECs~\cite{greiner2001exploring} effectively demonstrate phase transitions~\cite{lewenstein2012ultracold}.  The decay of correlations in a given phase may change form, e.g. from exponential to algebraic, but transitions at the critical point remain sharp.  In particular, in the Bose-Hubbard Hamiltonian (BHH), the most common model realized in cold atom quantum simulators, it takes only 5-10 sites before the quantum phase diagram begins to emerge~\cite{carr2010mesoscopic}.  In nuclear physics the study of such mesoscopic phase transitions is key due to the relatively small number of nucleons in a given nucleus, e.g. in nuclear shape transitions~\cite{iachello2001analytic}.  It is thus possible to create a finite-sized region of quantum matter in a quantum simulator set behind a barrier and observe the many-body quantum generalization of the original notion of tunneling, the quasi-bound or tunneling escape problem. As we will show, the quantum phase then determines the tunneling outcome.  Quantum simulators in which such experiments can be performed cover a wide range of architectures~\cite{altman2021quantum} in the quasi-1D context, including cold atoms in optical lattices, superconducting Josephson-Junction based circuits, and Rydberg chains, as all these systems can create mesoscopic quantum phases.

We focus here on cold atoms in optical lattices.  The 1D BHH has both a mean-field U(1) second order quantum phase transition and a Berzinskii-Kosterlitz-Thouless (BKT) or continuous quantum phase transition.  In Fig.~\ref{fig:sketch}(c) we show how such transitions appear for a mesoscopic system.  In Fig.~\ref{fig:sketch}(a)-(b) we show how the tunneling outcome is radically different between the wave-like, coherent, more semiclassical superfluid phase, and the atom-like, incoherent, interaction-induced Mott insulator phase.  However, density and phase do not suffice to predict the observed far-from-equilibrium dynamics.  As we will show, there are many other distinguishing features in the dynamical many-body quantum outcomes as observed in number fluctuations, entanglement, and two-point correlators.  For example, the Mott gap in the Mott insulator presents an extra barrier to overcome, leading to a surprising slowdown in quantum tunneling despite the stronger repulsive interactions in this quantum phase that would otherwise push the atoms more rapidly through the barrier.  Yet the peak of entanglement occurs much earlier, when only one quarter of the atoms have tunneled through as compared to one half for the superfluid phase.  Moreover, tunneling dynamics are non-exponential in both phases, in contrast to single-particle predictions, and even near-linear for the Mott insulating phase.  These and other surprising features are the result of quantum correlations, necessitating the introduction of a new quantum tunneling rate: the \emph{quantum fluctuation rate}.

This Article is outlined as follows. In Section~\ref{sec:Methods}, we show how the quantum phases of the
BHH, although slightly modified by the presence of the barrier, remain intact.  In Section~\ref{sec:Dynamics}, we present the results of our matrix-product-state simulations~\cite{jaschke2018open} on macroscopic quantum tunneling escape of a meta-stable
state into free space, calculating single-body observables like the number of atoms remaining in the trap, where we find distinct patterns of wave-like and atom-like tunneling for the superfluid and Mott-insulator interaction regimes, respectively.   In Sec~\ref{sec:beyond} we go beyond such traditional measures derived from the single-particle quantum tunneling picture, demonstrating that number fluctuations and von Neumann quantum entropy both show significant differences in the two quantum phases, and introduce the quantum fluctuation rate.  Finally, in Sec.~\ref{sec:correlations} we show how extremely different the tunneling dynamics of off-diagonal quantum correlations is in each case, highlighting that it is positive and negative quantum correlations that ultimately explain the difference in tunneling outcomes.   Our findings and conclusions are summarized in Section~\ref{sec:Conclusions}.

\section{Tunneling Initialization}
\label{sec:Methods}

In the following we describe the Bose-Hubbard Hamiltonian.  We describe the effects of mesoscopic confinement on the usual notion of the quantum phase.  Then we show that for sufficiently high barriers, scaled to interaction strength, the initial quantum state is well-confined and has a superfluid or Mott-insulating character on either side of the quantum critical point.  This sets up the problem for the study of tunneling dynamics in Sec.~\ref{sec:Dynamics}.

\subsection{The Bose-Hubbard Hamiltonian}
\label{ssec:BHH}

The BHH models cold bosonic atoms in
optical lattices in the tight-binding and lowest-band approximation, which is valid for typical atomic interaction strengths and a lattice potential energy several times the recoil energy or greater~\cite{bloch2012quantum}.  For weak interactions the BHH can alternately be considered as a discretization of the continuum field theory in the deep superfluid regime for long wavelength properties.  However, for strong interactions the BHH
undergoes a superfluid to Mott insulator quantum phase transition at a critical point $(J/U)_c=0.305$, where the BHH takes the form
\begin{equation}
\label{eq:BHH}
\hat{H} = -J\sum_{i=1}^{L-1}(\hat{b}_{i}^\dagger\hat{b}_{i+1}+\mathrm{h.c.})+\sum_{i=1}^L [\frac{U}{2}\hat{n}_i(\hat{n}_i-\hat{1})+V^{\mathrm{ext}}_i \hat{n}_{i}].
\end{equation}
The coefficients $J$ and $U$ are the hopping and on-site interaction energies,
respectively.  Hopping is often called ``tunneling'' but refers to a single-particle effect in which occupation of one lattice site tunnels to the next via an overlap integral between the site-local wavefunctions.   Here, we study in contrast \emph{macroscopic} quantum tunneling, referring to the collective tunneling of many atoms.  In order to study the macroscopic quantum escape problem, which can also be viewed as decay of a many-body quasi-bound state, we include an external, site-dependent, potential barrier as
$V^{\mathrm{ext}}_i$, where $i\in[1,L]$, and the total lattice size is $L$. Such a barrier can be realized e.g. by a tightly focused Gaussian beam on top of the lattice~\cite{potnis_interaction-assisted_2017}.  For the remainder of our study we work in hopping units, scaling all energies to the hopping energy, $J$, and time to
$\hbar/J$. For simplicity, we choose a square barrier of form
$V^{\mathrm{ext}}_i = h$ for $a<i<b$ with $a$ the well width and $w=b-a$ the barrier width, and $0$ otherwise.  Our BHH is stated in terms of finite atom number $N$, and therefore does not include a chemical potential term.  However, for the sketch in Fig.~\ref{fig:sketch}, $\mu$ may be taken as 

\begin{equation}
\label{eq:chempot}
    \mu(N)=\partial E / \partial N \simeq E(N+1)-E(N),
\end{equation} where $E=\langle \hat{H} \rangle$ with respect to the ground state or quantum phase.  The BKT phase transition occurs at the tip of the Mott lobe for commensurate filling, i.e. $N=L_{\mathrm{trap}}$, while the $U(1)$ mean field transition occurs as one transitions vertically through the phase diagram.  In our case, the latter translates into a noncommensurate filling created by subtracting atoms, as naturally occurs in the quantum escape process.

Our main solution method is matrix-product state (MPS) simulation, in particular time-evolving block decimation (TEBD) in our openMPS codes under imaginary time relaxation to obtain the initial state, and real-time propagation to determine tunneling dynamics.  Our usage of these open-source codes and convergence criteria are detailed thoroughly in~\cite{jaschke2018open} and have been established in prior works on the semiclassical limit to tunneling in~\cite{alcala_entangled_2017} and~\cite{zhao_macroscopic_2017}.  In summary, we converge in Schmidt truncation error, or error due to a only a finite number of elements retained in the reduced density matrix after a partial trace, and local dimension, allowing sufficient number fluctuations on-site.  These simulations are time-adaptive, as standard for MPS methods. Local dimension is converged from 4 to up to 8 atoms per site ($d=5$ to $d=9$ including the vacuum state of zero atoms on-site), while entanglement is converged with a Schmidt number of from $\chi=60$ to $\chi=200$.  In previous work much lower $\chi$ was required as superfluids are not highly entangled, but to capture Mott dynamics we needed to consider higher $\chi$ in this work.  All results are converged to much better than visible to the eye, and sufficient for the conclusions of this Article.  In particular, all curves and surfaces shown in figures have a maximal relative error of $10^{-2}$ at the longest times of $t=300$ to $500$ for the highest interaction strengths, where relative error is taken as $\varepsilon = |(f_1-f_2)/[2(f_1+f_2)]$ with $f_1$ and $f_2$ observables of increasing $\chi$, local dimension $d$, etc.  Within the key part of the dynamics at $t=0$ to $t=150$ we maintain a convergence of $\varepsilon \leq 10^{-4}$ in all observables.

\subsection{Mesoscopic Quantum Matter}

We first examine the effect of the barrier
on the ground-state parameter space for both commensurate and non-commensurate cases.
Our initial meta-stable state, localized inside the well of size $a=L_{\mathrm{trap}}
\lesssim N$, will be close to a commensurate filling, with a few atoms
penetrating into the barrier, Fig.~\ref{fig:Initial_State}. Under time
evolution, this state will tunnel into a near-continuum escape region of an extended lattice, where the
number of lattice sites far exceeds the number of escaping atoms.
The motivations behind this initial study
are twofold. First, we need to understand how the penetration of the tail of the many-body wavefunction into the barrier
will affect the Mott-to-Superfluid transition shown in Fig.~\ref{fig:sketch}.  Second, the statics will help clarify which values of
$U$ and $h$ to use in tandem. For example, if the barrier $h$ is too high,
then the wave function will tunnel too slowly to be observed within a reasonable
time scale for our simulations and for experiments, resulting in self-trapping, as observed also for the double well~\cite{albiez_direct_2005}. If $h$ is too low, the repulsive interactions in the trap will
overcome the barrier, the many-body wavefunction will spill classically over the top of the barrier, and no meta-stable states will exist.

\begin{figure}
    \subfloat{
        \includegraphics[width=0.77\linewidth]{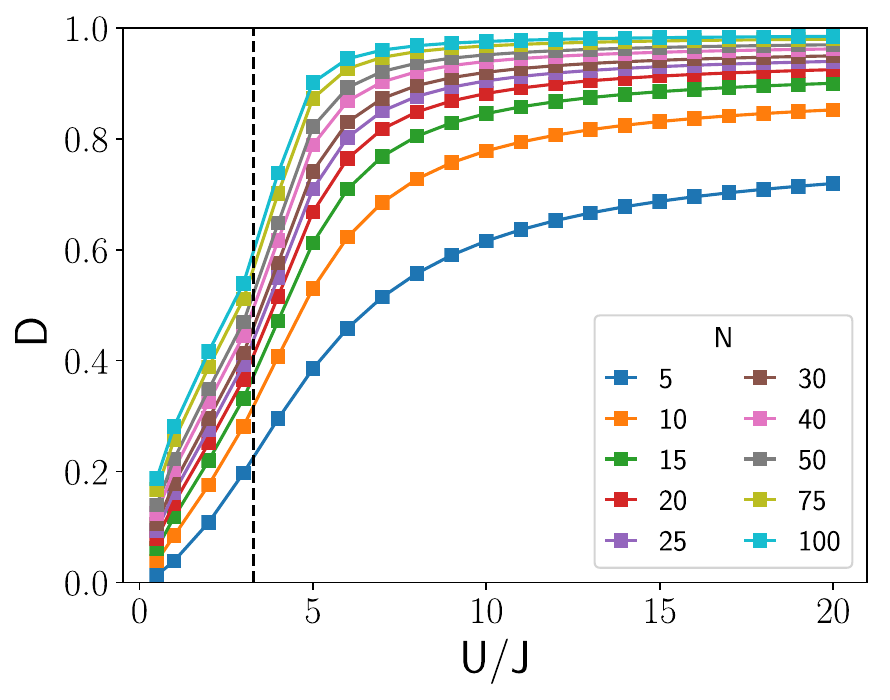}
    }
    \caption{\label{fig:Mott_to_SF_depletion}
        \emph{Finite size crossover in quantum depletion through the quantum phase transition.} Quantum depletion measures the fraction of atoms not in the dominant semiclassical mode in the superfluid or Bose-Einstein condensate (BEC) regime~\cite{bloch2012quantum}.  Shown are $N=5$ to $100$ atoms with commensurate filling $L=N$, as a
        function of $U/J$. As $N$ increases, the quantum depletion approaches an infinite system size limit.  High levels of quantum depletion approaching and beyond the quantum critical point at $(U/J)_c=1/0.305=3.28$~\cite{carrasquilla2013scaling} (black dashed vertical line) indicate semiclassical methods such as the JWKB and path integral approximations will fail~\cite{danshita2011superfluid}, necessitating our MPS approach to capture strong correlations.
    }
\end{figure}

We first consider a small uniform finite-size system, that is, the well only, without the barrier.  Figure~\ref{fig:Mott_to_SF_depletion} shows how quantum depletion, $D$, trends towards the infinite size limit for increasing $U/J$ and $N$, where 
\begin{equation}
\label{eq:depletion}
D = 1-\frac{\lambda_1} {\sum_{m=1}^{L}\lambda_m },
\end{equation}
with the eigenvalues, $\lambda_m$, determined from the single-particle density matrix, $\langle \hat{b}_i^{\dagger}\hat{b}_j\rangle$, and $\lambda_1$ the
largest eigenvalue.  Because a many-body treatment of our meta-stable state neccessarily has a
finite number of atoms, Fig.~\ref{fig:Mott_to_SF_depletion} roughly outlines how ``Mott-like'' or
``superfluid-like'' a finite commensurate filled system will behave. The true
BKT phase transition occurs at $(U/J)_{c} \approx 3.28$~\cite{carrasquilla2013scaling}. Even for just 5 or 10 sites, the quantum depletion rises rapidly as the theoretical quantum critical point from the infinite size extrapolation is crossed.  The quantum critical point for finite-size systems is often taken as the point of inflection in this curve~\cite{carr2010mesoscopic}; however, for simplicity, it suffices to refer to the theoretical infinite size limit for the rest of our paper, as we will test values of $U/J$ well to the left and right of the vertical line shown in Fig.~\ref{fig:Mott_to_SF_depletion}.

\begin{figure}
    \subfloat{
        \includegraphics[width=0.97\linewidth]{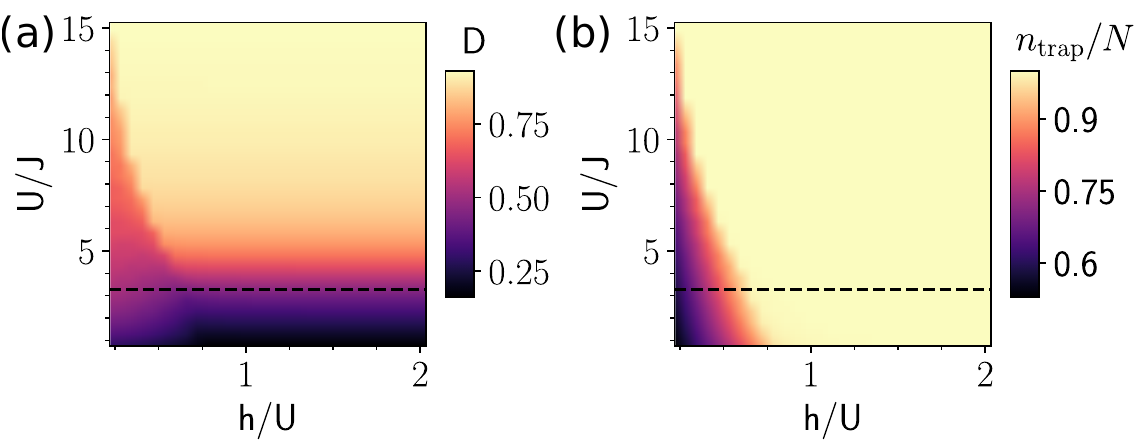}
    }
    \caption{\label{fig:mott_to_superfluid_barrier}
         \emph{Effect of the barrier on the quantum phase transition.}  
        (a) Quantum depletion and (b) number of trapped atoms as a
        function of interaction strength $U/J$ and barrier height $h/U$. The critical point for the quantum phase transition is indicated with a dashed black line.  In the Mott regime the depletion is only decreased by sufficiently small barriers $h/U \lesssim 0.5$, indicating the quantum phase is maintained. Likewise, interactions, although repulsive, prevent penetration into the barrier.
    }
\end{figure}

\subsection{Effects of the Barrier on the Quantum Phase}
\label{ssec:barrier}

Next, we consider the effects of the finite barrier.  In this Article, we study escape dynamics from a confinement area of size $L_\mathrm{trap}$ behind a narrow barrier into a quasi-continuum escape region, i.e., in systems with $L >> N$. Previous research,
analyzing double-well dynamics, looked at highly discrete systems with $L \sim 2 N$. In fact, a frequent approximation is the two-mode approximation or Lipkin-Meshkov-Glick model~\cite{lipkin1965validity}, which assumes just one discrete state on each side of the barrier~\cite{albiez_direct_2005,dounas2007ultracold}.  The
much larger lattice in this Article alters the ground state regimes, introducing
restrictions to achieve sufficient containment in the trap.
To allow observation of meta-stable quantum tunneling into a quasi-continuous free space, we
require a barrier that is balanced between being large enough to trap the
atoms, and sufficiently small to allow for tunneling rates on a reasonable
time scale. Furthermore, the interaction strength, $U/J$, must span
superfluid- and Mott-dominated regimes.  

To explore such questions, we first determine the ground state with TEBD for an initial very wide barrier beginning at $i=a$ and ending at $i=b=L$, i.e., covering the whole escape region.  This is the initial state, explored in Fig.~\ref{fig:Mott_to_SF_depletion}.  Dynamics begins in Fig.~\ref{fig:Lattice_heatmap} when we abruptly reduce $b$ to $a+w\ll L$, where $w$ is the barrier width, with the remaining escape region from $b$ to $L$ a quasi-continuum.  For dynamics, we often take $L=500$ or more to avoid reflections in the escape region over the time scale of the simulation.  

To illustrate maintenance of the quantum phase and penetration of the wavefunction into the barrier in Fig.~\ref{fig:Mott_to_SF_depletion}, we chose $N=25$, $a=25$, and $b=L=100$ for illustration purposes.  The quantum depletion $D$ and average scaled number of trapped atoms $n_{\mathrm{trap}}/N\braket{\hat{n}_{\mathrm{trap}}}/N$ both show a clear boundary as a function of interaction strength $U/J$ and barrier height $h/U$.  We choose to scale barrier height to $U$ rather than $J$ because in a semiclassical picture the effective barrier height for tunneling~\cite{zhao_macroscopic_2017} is $h-\mu\simeq h-U*1$, since $\mu\simeq U * n_\mathrm{trap}/L_\mathrm{trap}\simeq U * 1$.  Although a semiclassical picture proves insufficient for the Mott regime in particular, this is a good starting point as a baseline.  As long as the barrier is not too low with respect to the interaction strength, Fig.~\ref{fig:Mott_to_SF_depletion}(a) shows that the quantum phase is well maintained, while Fig.~\ref{fig:Mott_to_SF_depletion}(b) show the initial penetration of the wavefunction into the barrier is small.  

We explored from $N=100$ to $N=5$, and these effects persist for $N=L_\mathrm{trap}=a$ throughout this regime, although with slightly less sharp boundaries in Fig.~\ref{fig:Mott_to_SF_depletion} for smaller $N$.  These results are consistent with the mesoscopic quantum phase transition conclusions in~\cite{carr2010mesoscopic}, and show the presence of the boundary, as long as not too low, maintains the quantum phase even for small regions of quantum matter.

\section{Coherent Superfluid vs. Incoherent Mott Tunneling Dynamics}
\label{sec:Dynamics}

\begin{figure*}
	\subfloat{
		\includegraphics[width=0.70\linewidth]{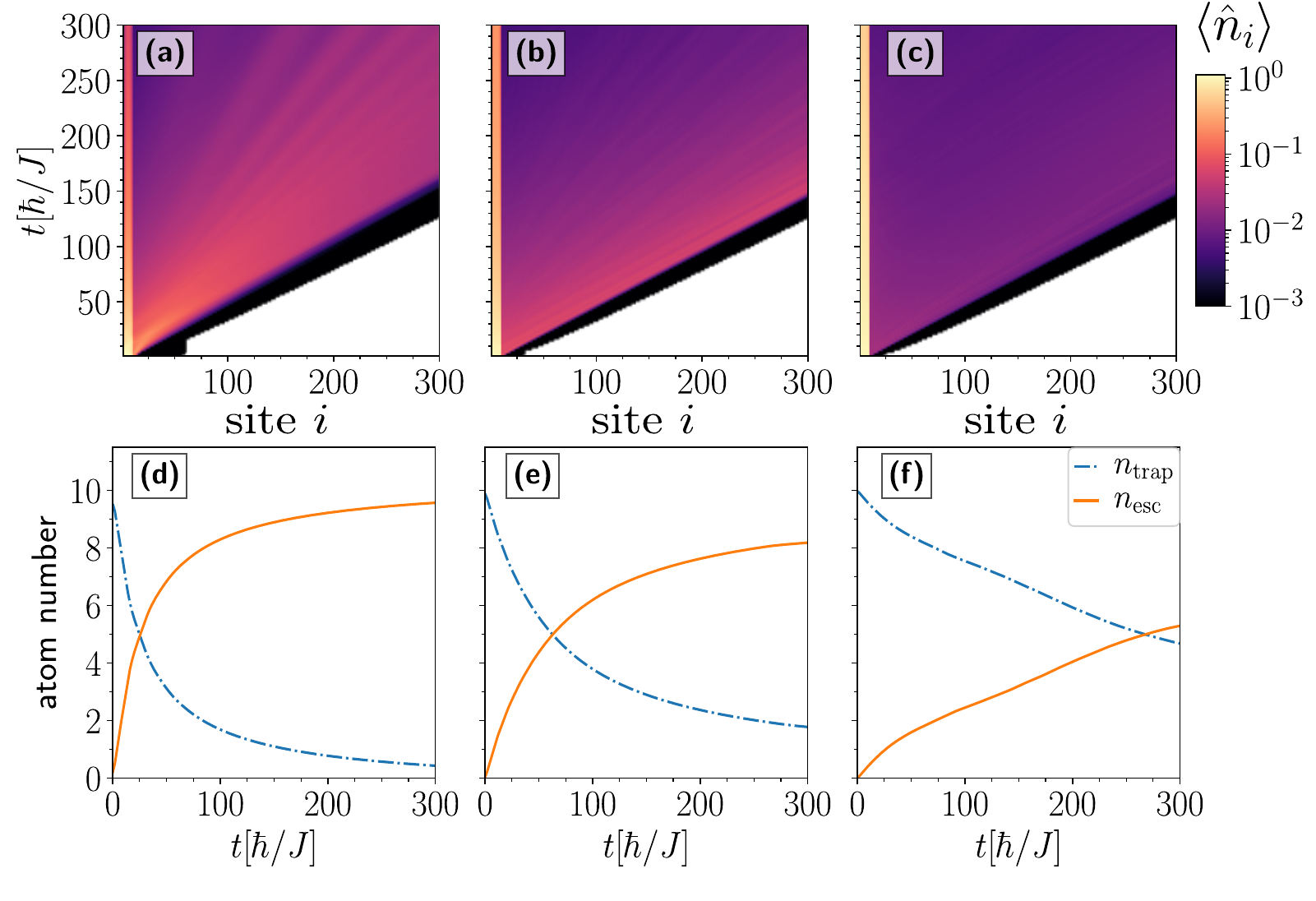}
	}
    \caption{\label{fig:Lattice_heatmap}
        \textit{Wave-like vs. atom-like quantum tunneling escape dynamics.}
        (a)-(c) Space-time evolution of average on-site occupation number $\braket{\hat{n}_{i}}$.  (d)-(f) Evolution of total atom number in the trap,
        $\braket{\hat{n}_\mathrm{trap}}$ and in the escape region,
        $\braket{\hat{n}_\mathrm{esc}}$.  Interactions increase from left to right: (a,d) weakly interacting superfluid regime, $U=1.0$; (b,e) critical regime, $U=3.0$; and
        (c,f) strongly interacting Mott insulator regime, $U=6.0$; all with $N=10$ and $h/U=1.0$.
        Weaker interactions show strong wave-like interference patterns, as one expects for a superfluid.  The Mott-insulator, despite being a ground state property, persists in the lack of interference in the incoherent escape dynamics, generating a much more atom-like behavior, albeit with a much slower rate influenced by the Mott gap.  The Mott gap creates an effective additional barrier to overcome, as one can observe in the slower decrease of atoms remaining in the trap in (f).}
\end{figure*}

After initializing the quasi-bound state as described in Sec.~\ref{ssec:barrier} via imaginary time propagation with TEBD, we drop the barrier except for a narrow delta-function-like remnant, creating a thin barrier through which the many-body quantum wavefunction can tunnel through on experimental timescales.  The rationale for such a thin barrier is key to making experiments work, and is detailed experimentally in~\cite{potnis_interaction-assisted_2017} and theoretically in~\cite{zhao_macroscopic_2017,alcala2018macroscopic}.  Thus the new barrier in
Eq.~\ref{eq:BHH} takes the form $V_{i} = h \delta_{i,N}$,
where $\delta_{i,N}$ is the Kronecker delta set so that the filling factor, or average occupation per site, is initially very close to 1, or commensurate.  

Thus, at the start of the dynamics, $t=0$, the wave function
is in a meta-stable state, able to escape into a quasi-continuum escape region. The hard wall at the end of lattice, at $i=L$, is taken sufficiently far that any reflected atoms do not interfere with the dynamics near the barrier, and typically chosen at $L=300$ to 500.  We proceed to propagate in real time with TEBD.  Figure~\ref{fig:Lattice_heatmap}(a-c) shows a space-time heatmap of the average on-site
atom number, $\braket{\hat{n}_{i}}$, with time along the vertical axis and
lattice site along the horizontal axis. From left to right are shown increasing interaction strength from an initial weakly-interacting superfluid ($U/J=1.0$ to a near-critical system $U/J=3.0$ to an initial strongly-interacting Mott insulator ($U/J=6.0$).  For the initial superfluid state, during the first $50$ time steps in Fig.~\ref{fig:Lattice_heatmap}(a), the escaped wave function stays together,
before fanning out into an interference-like pattern, with each anti-node
covering upwards of $10$ lattice sites, starting around $t \approx 90$ and
sites $i \ge 50$. These patterns have been called ``blips'' in semiclassical studies~\cite{dekel2007temporal}.  Such blips appear for both attractive and repulsive interactions, and even in the non-interacting or single-particle case, and are therefore due to interference phenomena obtainable with the Feynman propagator~\cite{carr2004spontaneous}.  In contrast, for initial critical and strongly-interacting regimes, Fig.~\ref{fig:Lattice_heatmap}(b-c), the wave-like interference phenomena disappear.  Instead, there are only weakly distinguishable and narrow streaks, immediately after the atoms start escaping.  Each streak is very narrow, and does not show a regular interference pattern.  Note that in all regimes the black line is a result of the well-known Lieb-Robinson bound~\cite{lieb1972finite} or ``quantum speed limit'' for Eq.~\eqref{eq:BHH}.

The bottom row of Fig~.\ref{fig:Lattice_heatmap} shows the number of atoms remaining in the trap, $n_\mathrm{trap}$ and the number that have escaped into the quasi-continuum, $n_\mathrm{esc}$. The number of atoms under the narrow barrier is always much less than 1, and is not shown.  For the superfluid regime it was previously demonstrated~\cite{zhao_macroscopic_2017,alcala_entangled_2017} that stronger
repulsive interactions, $U$, in general cause faster escape for a given barrier height,
$h \equiv \mathrm{const.}$.  This is because in the semiclassical limit repulsive interactions lead to an effective nonlinear term which push the tail into the barrier.  The dependence of rate on interaction strength as a function of $n_\mathrm{trap}$ is somewhat subtle, and in fact the rate very slightly decreases in a small region near the point of spilling classically over the barrier due to deformation of the barrier by the mean field or nonlinearity~\cite{alcala2018macroscopic}.  

However, consideration of the critical to strongly-interacting regime and solution with a fully entangled dynamical method as we perform here with TEBD shows a massive decrease in the tunneling rate, as observed in the bottom row of Fig~.\ref{fig:Lattice_heatmap}.  A key feature of the Mott insulator is the Mott gap, $\Delta=2U$.  This is the energetic barrier required to move one atom by one site, as evident in Eq.~\eqref{eq:BHH} for $U\gg J$.  In order for tunneling to occur in a Mott insulator atoms have to hop one site at a time, rather than all together and collectively as in the superfluid limit, and thus the Mott gap must be overcome.  Especially early in the tunneling process where the initial state has a Mott gap due to initial commensurate filling, here of 10 atoms on 10 sites, the Mott gap thus presents an additional barrier that must be overcome, decreasing the tunneling rate significantly.   This effect is sketched in Fig.~\ref{fig:sketch} qualitatively and born out here in dynamical simulations.  We observe the same kinds of slow-down effect for 25 atoms on 25 sites in critical and strongly interacting regimes (simulations not shown).

Loosely speaking, we may quantify this transition from superfluid to Mott-insulating regimes as wave-like to atom-like.  In the wave-like limit a semiclassical theory provides guidance, and we see a clear and regular pattern of interference fringes.  In the atom-like limit tunneling is dominated by atom-like hops influenced by the Mott gap.  We remind the reader that Mott insulator refers to the resistance of the quantum state to atom flow, or current.  Escaping through the barrier is analogous to water flowing through a break in a dam, or current through a weak point in a barrier as in a Josephson Junction.  The superfluid flows in the Josephson regime.  Placing a Mott insulator behind the barrier greatly reduces the ability of the atoms to rapidly flow.

\begin{figure}
    \subfloat{
        \includegraphics[width=0.97\linewidth]{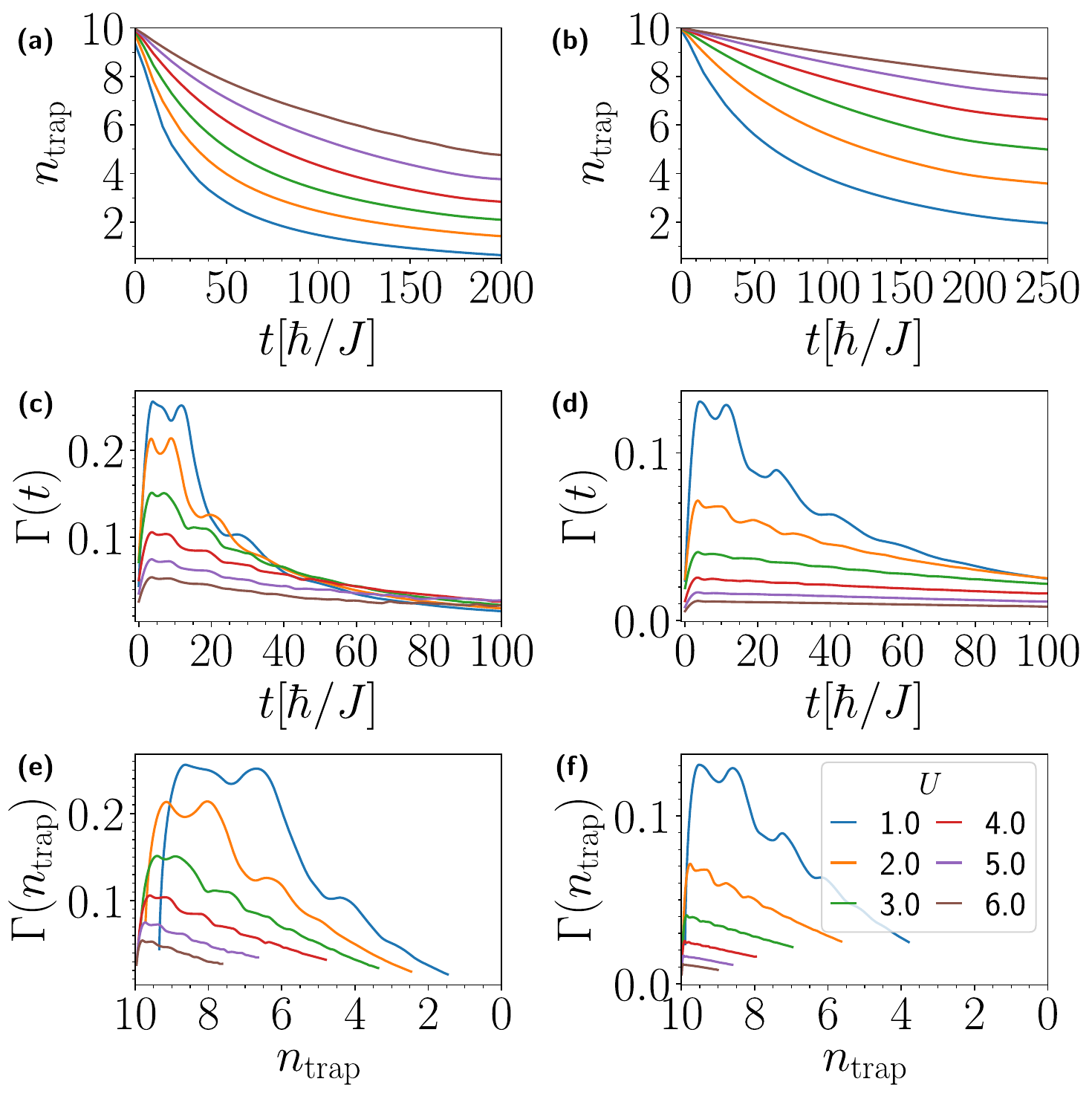}
    }
    \caption{\label{fig:n_trap_and_n_trap_derivatives_h_over_U}
        \textit{Non-exponential tunneling rates from superfluid to Mott-insulating regimes.}
        (a)-(b) Number of trapped atoms $n_\mathrm{trap}$ for increasing interaction strengths from weakly interacting to critical to strongly interacting regimes show a rapid slow-down in tunneling, in contrast to semiclassical predictions.    (c)-(d) Time-dependent tunneling rates $\Gamma(t)$ and  (e)-(f) number-dependent tunneling rates $\Gamma(n_\mathrm{trap})$ show a highly non-exponential behavior, with wave-like interference phenomena for weak interactions and an order-of magnitude difference in rates as a function of interactions.  The 
        low barrier case of $h/U=0.9$ (left column) shows more persistent wave-like interference patterns as compared to the high barrier case of $h/U=2.0$ due to weaker and therefore less commensurate confinement of the initial state.  The initial Mott insulator is affected more by the Mott gap when more strongly initially confined.}
\end{figure}

We already know that the decay curve of $n_\mathrm{trap}$ is non-exponential~\cite{potnis_interaction-assisted_2017,zhao_macroscopic_2017} even in the weakly interacting superfluid regime.  This is interpreted as being due to the single-particle energy (equivalent to a chemical potential) dropping relative to the barrier height.  However, here we see that for the Mott insulator the distortion from the well-known single-particle exponential form is much more extreme.  To examine this question more closely, in Fig.~\ref{fig:n_trap_and_n_trap_derivatives_h_over_U}(a)-(b) we show the dependence of $n_\mathrm{trap}$ on both the barrier height $h/U$ and the interaction strength $U/J$.  As we described in Sec.~\ref{sec:Methods}, the barrier height is scaled with interaction to keep the effective barrier height at the same level, as the effective chemical potential in Eq.~\eqref{eq:chempot} scales with $U$ and sets the single-particle tunneling energy in the presence of the trapped many-body wavefunction.  Although the slow-down in the rate for stronger interactions is easily apparent, the time-dependent rate $\Gamma(t)=dn_\mathrm{trap}/dt$ clarifies the extreme difference in the non-exponential behavior beyond wave-like or atom-like classifications.  For single-particle quantum tunneling the rate equation takes the form $dn/dt = -\Gamma n$, with $\Gamma$ constant.  Here, whether plotted for $\Gamma(t)$ in Fig.~\ref{fig:n_trap_and_n_trap_derivatives_h_over_U}(c)-(d) or $\Gamma(n_\mathrm{trap})$ in Fig.~\ref{fig:n_trap_and_n_trap_derivatives_h_over_U}(e)-(f), $\Gamma$ is very clearly non-constant and therefore non-exponential.  The rates are calculated from numerical derivatives on our data in Fig.~\ref{fig:n_trap_and_n_trap_derivatives_h_over_U}(a)-(b) using Python's SciPy interpolating function.

For weak interactions at $U=1.0$ the wave-like interference effects are apparent in oscillations in the rates, and the rates are initially rapid, then slow down. However, at the critical point of $U=3.0$ and beyond into the strongly interacting Mott insulator regime of $U=6.0$, the rates are an order of magnitude smaller.  They at first increase rapidly on a very short time scale, then decrease linearly as a function of time.  In the case of $\Gamma(n_\mathrm{trap})$ one can see very small oscillations in the rate corresponding to the faint streaks seen in Fig.~\ref{fig:Lattice_heatmap}(c).  The narrow width of these indicates single atom effects.  The Mott gap prevents more than one atom leaving the system at a time, or indeed any kind of collective escape, as for two atoms to act together they must overcome the Mott gap of $2U$ over and above the trap barrier.  Lower barriers (left column of Fig.~\ref{fig:n_trap_and_n_trap_derivatives_h_over_U}) allow wave-like interference effects to persist to higher interaction strengths as compared to higher barriers (right column of Fig.Fig.~\ref{fig:n_trap_and_n_trap_derivatives_h_over_U}).  This is due to higher barriers creating a more commensurate initial state with strong confinement, as also observed in Fig.~\ref{fig:Mott_to_SF_depletion}.

\section{Number Fluctuations, Entropy, and a New Rate to Characterize Macroscopic Quantum Tunneling}
\label{sec:beyond}

So far we have looked at how the typical observables from the single-particle quantum tunneling escape problem are modified by interactions and an initial quantum phase.  We observed wave-like and atom-like dynamical in the space-time dependence of the number density and a highly non-exponential decay rate.  However, in many-body quantum systems we can also measure new quantities which provide new information not relevant to a single-particle picture.  

In the primarily mean field or semi-classical picture of macroscopic quantum tunneling explored in many weakly-interacting or statistically driven scenarios prior to this Article~\cite{albiez_direct_2005,potnis_interaction-assisted_2017}, number fluctuations were necessarily zero, as the mean field approximation neglects these.  An outstanding question has thus been how number fluctuations affect tunneling dynamics, over and above the semiclassical limit.  Here we can track their evolution explicitly with TEBD, and determine a new quantum tunneling rate, the \emph{quantum fluctuation rate}, which clearly demarcates the boundary between superfluid and Mott insulating phases.  

An understanding of number fluctuations is also important because bipartite entanglement measures like the bond entropy between the trapped and escaped atoms has been shown to be driven by local fluctuations in a globally conserved quantity~\cite{song2012bipartite}. Since total atom number is conserved in our TEBD simulations, we can consider first number fluctuations in~\ref{sec:fluctuations}, then explore the generated quantum entropy in Sec.~\ref{sec:entropy}.

\subsection{Number Fluctuations}
\label{sec:fluctuations}

Number fluctuations can be defined on a single site as
\begin{equation}
    \Delta (n_i)^2\equiv\braket{\Delta (\hat{n}_i)^2}=  \braket{\hat{n}_i^2} - \braket{\hat{n}_i}^2
\end{equation}
or between trap and escaped region as
\begin{equation}
    \Delta (n_\mathrm{trap})^2\equiv\braket{\Delta (\hat{n}_\mathrm{trap})^2}=  \braket{\hat{n}_\mathrm{trap}^2} - \braket{\hat{n}_\mathrm{trap}}^2
\label{eq:Fluc_rel}
\end{equation}
with $\hat{n}_\mathrm{trap}=\sum_{i=1}^{\ell}\hat{n}_i$ the sum over number operators for all atoms remaining in the trap.  

\begin{figure}
    \subfloat{
        \includegraphics[width=0.77\linewidth]{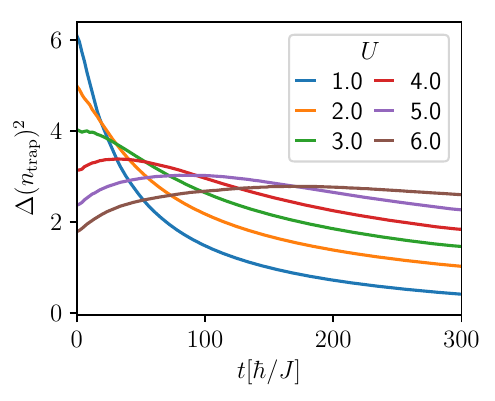}
    }
    \caption{\label{fig:number_fluc_time}
        \textit{Time evolution of number fluctuations in trap.}
        Many previous models of macroscopic quantum tunneling focused on the semiclassical or mean field limit in which number fluctuations are assumed to be zero.  Here we show that they depend strongly on the initial quantum phase, with a monotonic decrease in the weakly interacting superfluid regime and a non-monotonic increase followed by a decrease in the Mott insulating regime.  The transition occurs around the critical point.
        }
\end{figure}

In Fig.~\ref{fig:number_fluc_time} is shown the time evolution of the number fluctuations in the trap for a barrier of height $h/U=1.0$.  Number fluctuations start at 6, or $6/10=60\%$, for $U/J=1.0$.  They then rapidly decrease to zero in the superfluid regime as the tunneling proceeds.  As repulsive interactions increase, the starting level of number fluctuations is lower, and the decrease is slower, but the behavior is still monotonic.  However, starting in the critical region at $U/J=3.0$, the time evolution changes character, turning from concave to convex, and for stronger interactions into the Mott insulating regime the number fluctuations become non-monotonic.  They at first rise, then decay slowly, with a time scale that grows as the interactions are made stronger.  

\begin{figure}
    \subfloat{
        \includegraphics[width=0.77\linewidth]{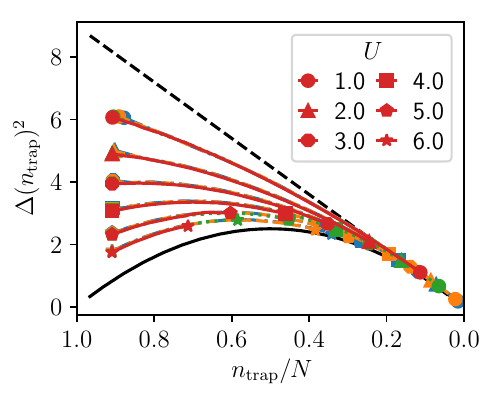}
    }
    \caption{\label{fig:number_fluc_combined}
        \textit{Number fluctuations as a function of the remaining trapped atoms.}  Number fluctuations dynamics depend principally in interaction strength $U/J$ ($U=1$ to $U=6$, symbols in key).  Different barrier heights of $h/U = 0.9, 1.0, 1.5, 2.0$ (blue, orange, green, and red) collapse onto nearly the same curves.  The extreme $U/J = 0$ and $U/J = \infty$ limits are shown as dashed black and solid black curves, respectively, clarifying the emergence of non-monotonicity in the time evolution in Fig.~\ref{fig:number_fluc_time}.
        }
\end{figure}

To help interpret this distinct non-monotonicity of the number fluctuations in the Mott insulating phase in Fig.~\ref{fig:number_fluc_combined} we plot the number fluctuations as a function of the number of remaining atoms in the trap, normalized to the total atom number $N$. The dashed curve shows the extreme case of $U/J \to 0$, while the solid curve shows the case of $U/J \to \infty$ -- both can be calculated straightforwardly from perturbation theory.  In the $U/J \to \infty$ limit, the number fluctuations start at zero because the system is in a Fock state of one atom per site.  As the tunneling proceeds, one particle at a time passes through the barrier, increasing number fluctuations as more holes appear in the system, since atoms can hop both left and right.  When half the atoms have left the trap, fluctuations are maximal, and then decrease as the number of arrangements of $n_\mathrm{trap}$ atoms on $\ell$ sites decreases.

In Fig.~\ref{fig:number_fluc_combined}, we observe that it is at the critical point where the transition between monotonic and non-monotonic behavior occurs.  In Fig.~\ref{fig:number_fluc_time} we showed only the barrier heigh $h/U=1.0$; here we show that all barrier heights nearly collapse onto the same curves, and that number fluctuations depend mainly on interaction strength.  

\begin{figure}
    \subfloat{
        \includegraphics[width=0.77\linewidth]{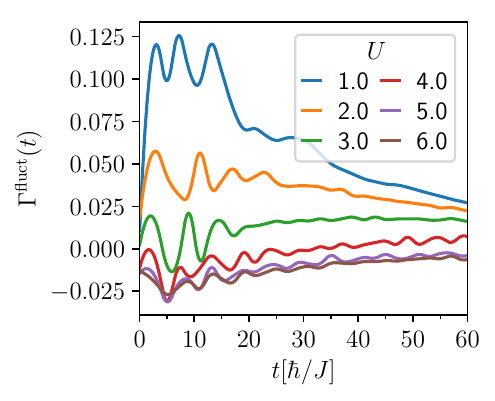}
    }
    \caption{\label{fig:number_fluc_derivative}
        \textit{Rate of number fluctuations in the trap.}  In single-particle quantum tunneling there is only one rate, proportional to the survival probability of an escaping atom.  Here we show a new rate for the many-body tunneling problem, the \emph{quantum fluctuation rate}, exhibiting a clear boundary between quantum phases.  Below the critical point the superfluid has a positive quantum fluctuation rate.  Above the critical point the Mott insulator has an intial negative rate, which only later become positive as seen in Figs.~\ref{fig:number_fluc_time}-~\ref{fig:number_fluc_combined}.  Initial oscillations on the time scale of $t J / \hbar = 0$ to 20 correspond to internal reflections within the trap  of size $L_\mathrm{trap}=10$.
        }
\end{figure}

Although the evolution of the number fluctuations appears smooth, in fact on shorter time scales it is oscillatory.  Early time evolution of even single-particle quantum tunneling is known to be non-exponential~\cite{wilkinson1997experimental}, due to a waiting period for tunneling to begin, that is, the exponential decay of single-particle tunneling does not turn on instantaneously.  In Sec.~\ref{sec:Dynamics} we demonstrated non-constant rates in the average number over and above  single-particle expectations.  To complete our study of number fluctuation dynamics, we consider the rate of change of number fluctuations, 
\begin{equation}
    \Gamma^{\mathrm{fluct}}(t)\equiv -\frac{d}{dt}[\Delta(n_\mathrm{trap})^2].
\end{equation}
In Figure~\ref{fig:number_fluc_derivative} we show that initial oscillations in the number fluctuations occur mainly from $t=0$ to $t=20$.  For a trap of size $L_\mathrm{trap}=10$, this is the time for excitations at the barrier edge to reflect back through the trap and interfere in the escape process.  Such oscillations are much weaker for the Mott insulator as they must flow over the top in a superfluid ``skin'' as seen also in the well-known wedding cake structure in trapped BHH systems~\cite{jreissaty2011expansion}.  Once these trapped oscillations created by the initial state escape, the rate is positive and rapidly decreases for the superfluid.  However, the Mott insulator has an initial negative quantum fluctuation rate, which only later becomes positive.  The critical point determines where the quantum fluctuation rate passes from positive to negative as interactions are increased.  

We emphasize the number fluctuations can be determined experimentally in BECs and cold atoms in optical lattices by subtracting the ensemble average of many density measurements from each individual density measurement.  The ensemble over the resulting images is related to the average fluctuations~\cite{altman2004probing}, from which the average quantum fluctuation rate can be determined.

\subsection{Quantum Entropy}
\label{sec:entropy}

\begin{figure}
    \subfloat{
        \includegraphics[width=0.97\linewidth]{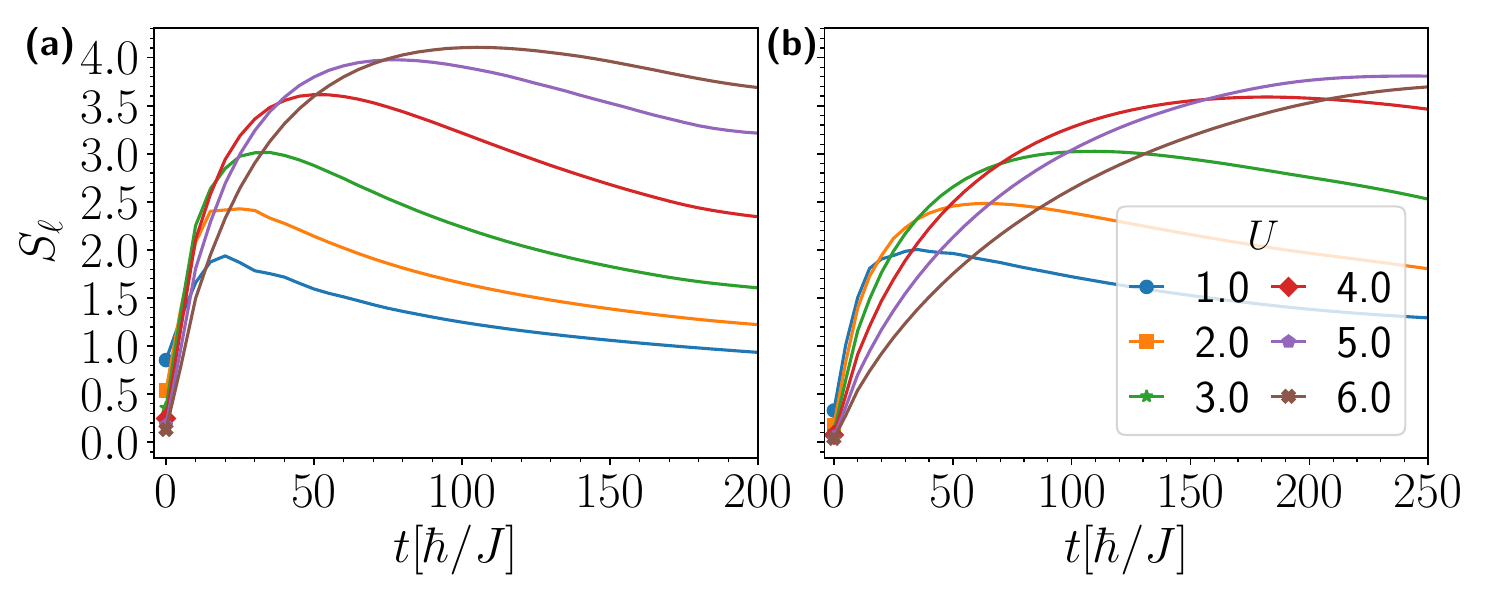}
    }
    \caption{\label{fig:entropy_time_over_h_U}
        \textit{Trap quantum entropy dynamics for superfluid and Mott insulator.} 
        The von Neumann entropy of entanglement between trapped and escaped atoms, $S_{\ell}$, as a function of time for (a) lower barrier height $h/U=1.0$ and (b) a higher barrier of $h/U=2.0$. The entropy shows 
        two distinct regimes during the tunneling escape dynamics: rapid rise to a peak value which depends on interactions, and a slow decay.  The rise time is slower for higher barriers as a few atoms must tunnel for entropy to build up.  The maximal entropy is about twice as high for the strongly interacting regime of an initial Mott insulator behind the barrier.
        }
\end{figure}

There are many entropy measures in a quantum many-body system.  The most relevant one for macroscopic quantum tunneling is the entropy of entanglement generated by escaping atoms.  The resulting density matrix of the atoms remaining in the trap is
\begin{equation}
    \rho_{\ell}=\mathrm{Tr}_{j>\ell}(\rho),
\end{equation}
where $\ell$ is taken as the site at the right-most edge of the barrier, and $\rho=\ket{\psi}\bra{\psi}$ is the pure state density matrix formed from the complete time-dependent state.  The resulting mixed state  has an associated quantum entropy of the trap of
\begin{equation}
    S_\ell=-\mathrm{Tr}(\rho_\ell \ln \rho_\ell),
\end{equation}
which quantifies the lack of information about the remaining atoms due to the escaped atoms not being measured.  This is in fact a bond entropy, and is a well-known quantity characterizing the convergence of MPS methods~\cite{jaschke2018open} as well as area vs. volume law scaling,Page curves, and information scrambling~\cite{nakagawa2018universality}.  In our case the bond entropy grows but remains close enough to an area law to be simulatable, as seen in Fig.~\ref{fig:entropy_time_over_h_U}.  To see this, first observe that for $N$ atoms on $\ell$ sites without other truncations the size of the Hilbert space is $N + \ell - 1$ choose $N$, due to number conservation in the initial state.  Then the Hilbert space dimension $\Omega$ for $N=10$ and $\ell=11$ is $\Omega=184,756$.  The maximal entropy for a maximally mixed state is $S_{\ell}=\ln(\Omega)=12.13$.  As we observe in Fig.~\ref{fig:entropy_time_over_h_U}, the maximal entropy ranges from about 2 to about 4 in the weakly to strongly interacting regimes.  The bond entropy at other points in the system is lower.  Our maximal Schmidt number $\chi=200$ yields an entropy of $\ln(200)=5.29$.  This is another way to demonstrate convergence beyond the discussion in Sec.~\ref{ssec:BHH}.

The initial state at $t=0$ in Fig.~\ref{fig:entropy_time_over_h_U}(a) has a higher trap quantum entropy $S_\ell$ for weak interactions because the superfluid has a tail extending further into the barrier. This effect is suppressed by a higher barrier, shown in Fig.~\ref{fig:entropy_time_over_h_U}(b).  As the tunneling escape process ensues the initial superfluid and Mott states at first sight show similar dynamics. A rapid rise time is followed by a slow decay.  Both the rise time and the decay time are slower for stronger interactions, and seem to vary smoothly with $U/J$.  A higher barrier again differentiates these regimes more strongly, just like with the suppression of coherent interference patterns observed in Sec.~\ref{sec:Dynamics}.  This is because a higher barrier makes the initial state more strongly commensurate, and therefore a better Mott insulator in the strongly interacting limit.  

\begin{figure}
    \subfloat{
        \includegraphics[width=0.97\linewidth]{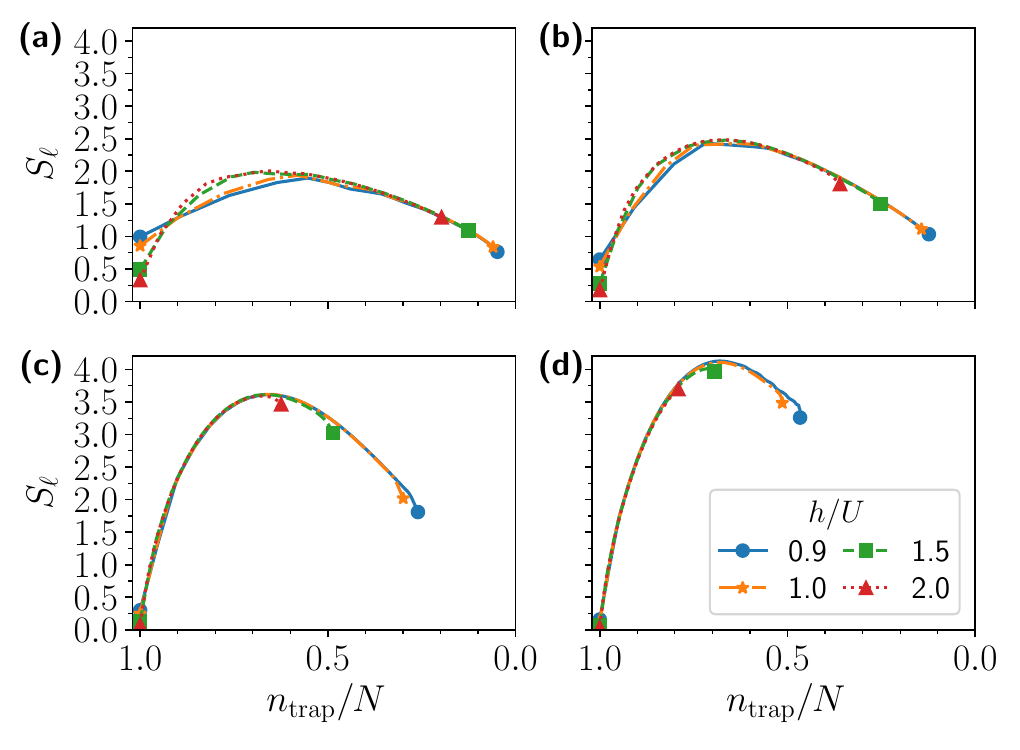}
    }
    \caption{\label{fig:entropy_ntrap_over_U}
        \textit{Trap quantum entropy dependence on the number of escaped atoms.} $S_\ell(n_\mathrm{trap}/N)$ for trap heights $h/U$=0.9, 1.0, 1.5, 2.0 and increasing interaction strength from
        (a) superfluid regime, $U=1$, (b) $U=2$, (c) $U=4$, to (d) the Mott insulator regime $U=6$, all for $N=10$.  The effects of the barrier are very weak in this view, whereas the effects of interactions are quite noticeable.  The maximal entropy of the superfluid is about half that of the Mott insulator.  In the superfluid case, the trap entropy maximizes when about half the atoms have tunneled, while in the Mott insulator case, maximization occurs when only about one quarter of the atoms have escaped.
        }
\end{figure}

However, there are in fact two key differences between the superfluid and Mott insulating regime.  First, the Mott insulator shows a maximal entropy which is about twice that of the superfluid, as observed in Fig.~\ref{fig:entropy_time_over_h_U}.  This maximum occurs slightly later in time, but considering how slowly the Mott insulator tunnels, it is important to ask not at what time the entropy maximum occurs, but for how many escaped atoms it occurs.  In Fig.~\ref{fig:entropy_ntrap_over_U} we plot the trap entropy as a function of the number of escaped atoms, scaled to the total number.  In this view, all barrier heights nearly collapse onto each other for a fixed interaction strength, except in the initial state in Fig.~\ref{fig:entropy_ntrap_over_U} where in the superfluid regime the initial penetration of the tail into the barrier is suppressed by higher barriers, therefore decreasing the initial entropy.  Thereafter the tunneling escape dynamics, as followed by the number of atoms tunneled collapses nearly all onto the same curve, and depends mainly on the interaction strength.  This brings us to the second key difference between the superfluid and Mott insulator regime.  In the superfluid case, the trap entropy maximizes when about half the atoms have tunneled, while in the Mott insulator case, maximization occurs when only about one quarter of the atoms have escaped.
Thus in the Mott insulating case it takes only a small fraction of the atoms to carry away the maximal amount of information and drive the remaining trapped atoms toward their maximum entanglement, as compared to the superfluid.

\section{Number Correlations in Macroscopic Quantum Tunneling}
\label{sec:correlations}

So far we have considered only local or scalar terms, such as the tunneling rate, the quantum fluctuation rate, and the quantum entropy in the trap during the escape process.  However, quantum phases are best characterized by second-order correlations~\cite{sachdev2007quantum,carr2010understanding}.  For the superfluid to Mott insulator transition, these take the form of number correlations.  The number fluctuations studied in Sec.~\ref{sec:fluctuations} and the quantum fluctuation rate considered only the local, diagonal part of the number correlations.  We now consider local simultaneous measurements of separated regions, or the off-diagonal part.  In Sec.~\ref{ssec:positive} we highlight the role of positive and negative correlations in the superfluid and Mott insulator regimes, and in Sec.~\ref{ssec:laser} we present preliminary evidence of an application using the barrier to control pulsed and continuous-wave correlations.

\subsection{Positive and negative correlations}
\label{ssec:positive}

\begin{figure}
    \subfloat{
        \includegraphics[width=0.97\linewidth]{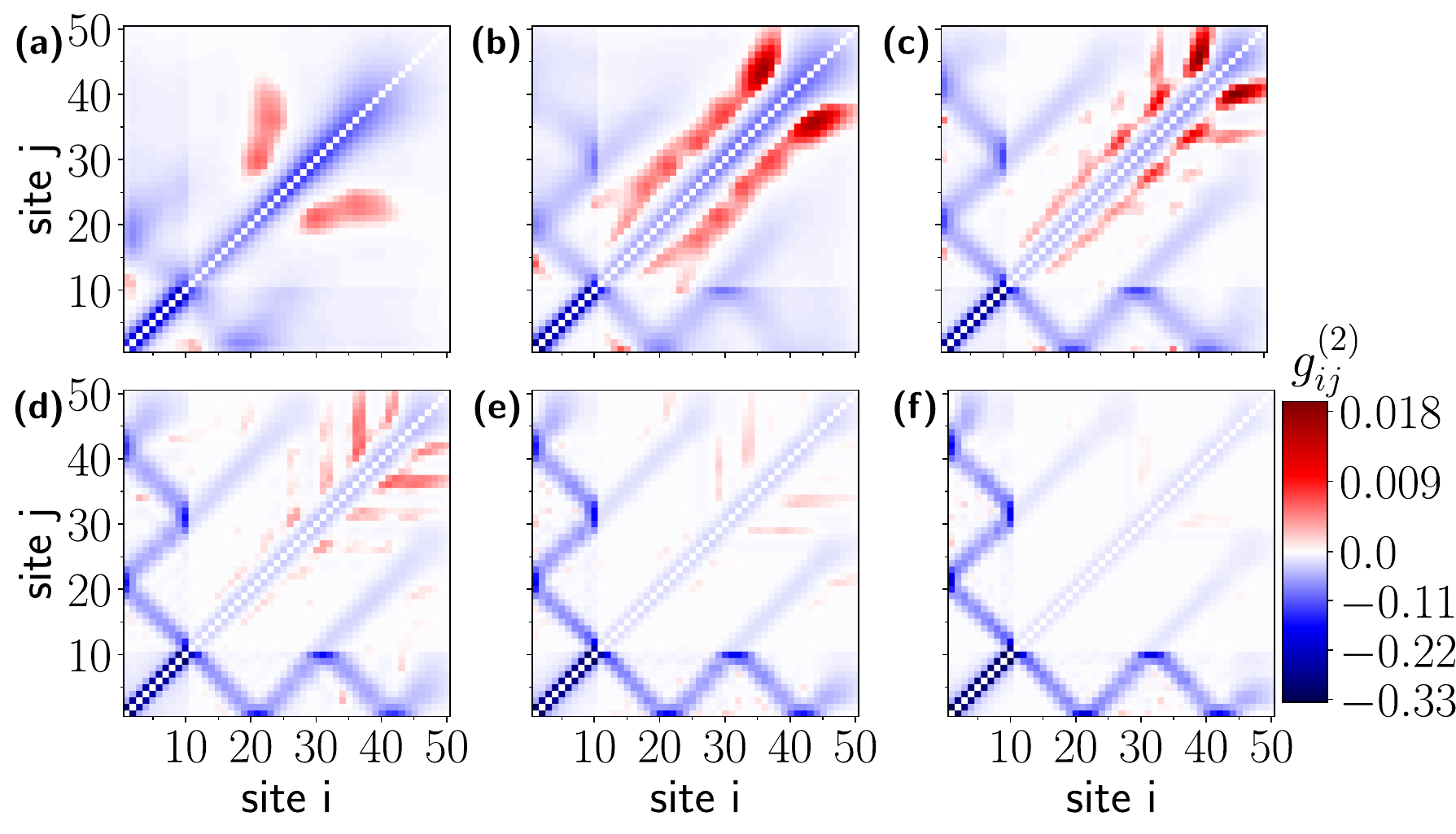}
    }
    \caption{\label{fig:g2_for_U}
        \textit{Off-diagonal number correlations at the onset of macroscopic quantum tunneling.}
        Two-point correlators at $t = 20$, $N=10$, $L_\mathrm{trap}=10$ for increasing interaction strength from superfluid regime (a) $U=1$ and (b) $U=2$ to the critical region (c) $U=3$ to (d) $U=4$ to the strongly interacting Mott insulating regime from (e) $U=5$ to (f) $U=6$.  Starting near the critical interaction strength, $U=3$, a multiple off-diagonal ``fork-like'' feature emerges and is persistent. Above $U=4$, the positive correlation regions become increasingly less pronounced.  Note negative (blue) and positive (red) correlations have separate scales to highlight the role of positive correlations more clearly.
    }
\end{figure}

In order to examine the question of the evolution of the quantum phase during the tunneling process, we therefore turn to the second order number correlator, given by
\begin{align}
\label{eq:g2}
g_{ij}^{(2)} & \equiv \frac{\biggl\langle \bigl(\hat{n}_i - \braket{\hat{n}_i} \bigr) \bigl(\hat{n}_j - \braket{\hat{n}_j} \bigr) \biggr\rangle}{\sqrt{\braket{\Delta n_i}^2} \sqrt{\braket{\Delta n_j}^2} } \\
& =\frac{\braket{\hat{n}_i \hat{n}_j} - \braket{\hat{n}_i} \braket{\hat{n}_j}}{\sqrt{\braket{\hat{n}_i^2} - \braket{\hat{n}_i}^2} \sqrt{\braket{\hat{n}_j^2} - \braket{\hat{n}_j}^2} }
\end{align}
Throughout our plots of $g_{ij}^{(2)}$, we subtract off the diagonal correlations as otherwise off-diagonal correlations can only be seen on an inconvenient log scale.  In Fig.~\ref{fig:g2_for_U} we show Eq.~\eqref{eq:g2} at the onset of macroscopic quantum tunneling highlighted in our study of the quantum fluctuation rate in Fig.~\ref{fig:number_fluc_derivative}.  For a trap of size $L_\mathrm{trap}=10$, at $t = 20$ in our units of $\hbar/J$, oscillations damp out as the internal reflections within the trap have had time to escape.  In all panels, the two pale horizontal and vertical lines at $i=11$ and $j=11$ indicate the presence of the barrier, dividing the plots into four distinct regions.  The lower-left $10\times 10$ region of $i,j \leq 10$ shows correlations within the atoms remaining in the trap.  The vertical-left region for $j\leq 10$ and $i > 11$ is equivalent to the lower right region with $i\Longleftrightarrow j$ corresponds to correlations between the trap and the escaped region.  Finally, the large upper-right region with $i,j > 11$ contains correlations purely within the escaped region.

We first observe that positive correlations are created in the escaped region for the initial superfluid regime in Fig.~\ref{fig:g2_for_U}(a)-(b).  This is despite the fact that only negative correlations show up within the trap.  Thus tunneling of a superfluid through a barrier creates positive correlations where none existed before.  We interpret this as due to bunching: atoms tend to tunnel together in clusters a few at characteristic scales of a 1-3 lattice sites, due to bosonic statistics.  A very small amount of positive correlation can also be observed between the sites 1 and 11, but this is due only to the internal reflections at time $t=20$.  Positive correlations extend into the critical region Fig.~\ref{fig:g2_for_U}(c), but quickly begin to fragment in Fig.~\ref{fig:g2_for_U}(d).  For the strongly interacting  Mott-insulator in Fig.~\ref{fig:g2_for_U}(f), they have disappeared entirely.

In contrast, negative correlations are most pronounced in the Mott insulating regime.  In Fig.~\ref{fig:g2_for_U}(e)-(f) a fork-like structure emerges.  The sharp lines in the structure emphasize the particle-like tunneling described in Sec.~\ref{sec:Dynamics}, here seen very clearly.  When a particle has tunneled into the escape region, it is subtracted from the trap region, leading to a clear negative correlation.  The slope of the lines in the fork is determined by our choice of units, and is just identical up to a sign throughout this region, $\pm 10$, for 10 sites traversed in 10 time units due to the trap-size of 10.  The reflections in the trap-escape region thus occur at 10, 20, 30, 40, etc.  This structure emerges in Fig.~\ref{fig:g2_for_U}(c)-(d) showing that the transition to particle-like tunneling occurs in the critical region. The transition is not completely sharp due to the meoscopic nature of the quantum phase transition as seen in Fig.~\ref{fig:sketch}(c).  For larger systems we expect it to be much sharper, but as many quantum simulators outside the field of cold atoms are expected to have about 10 quantum components (qubits, qudits, etc.) on the NISQ computing time scale~\cite{altman2021quantum}, we focus on the features already readily apparent at mesoscopic scale.

\begin{figure}
    \subfloat{
        \includegraphics[width=0.97\linewidth]{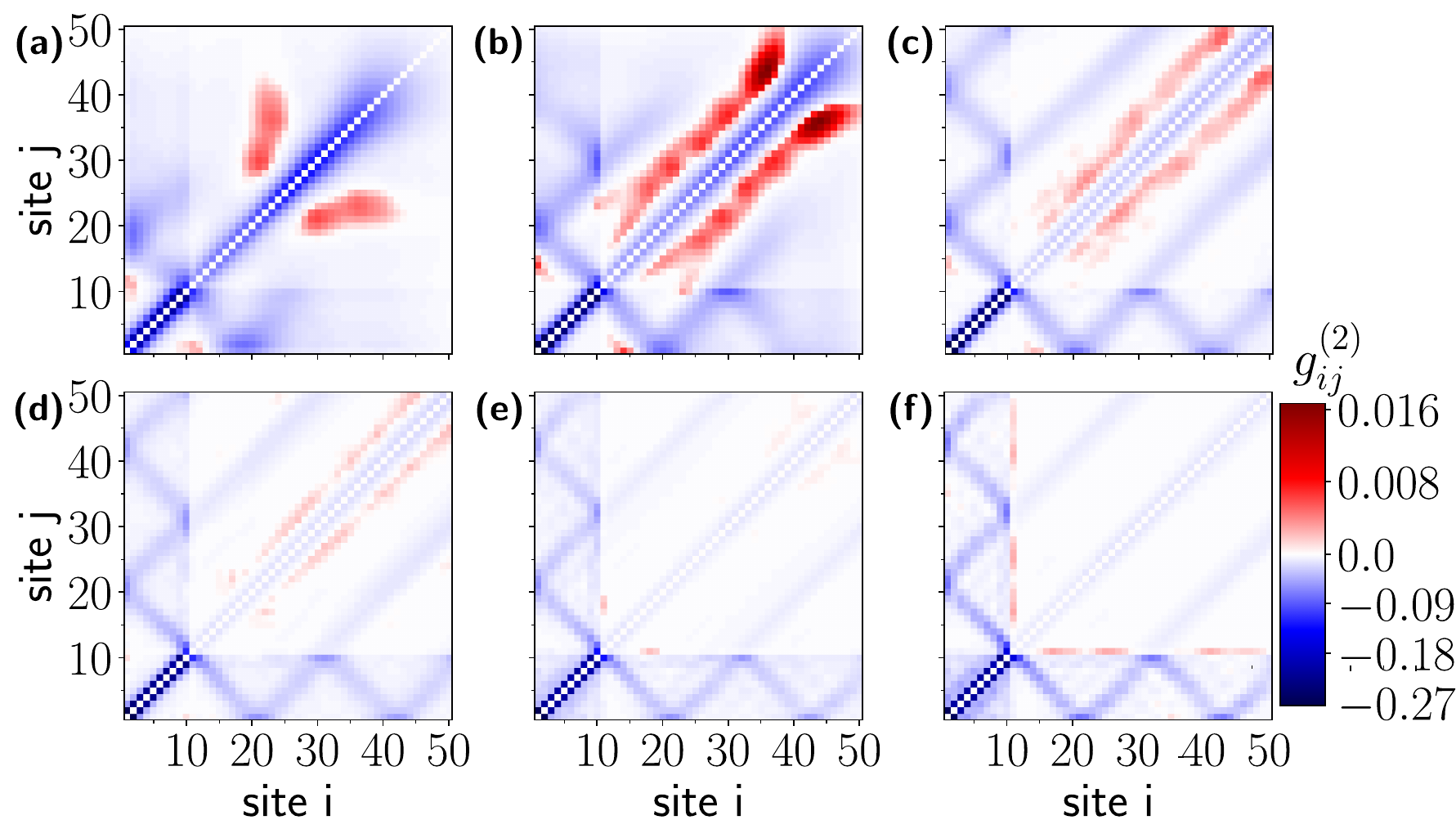}
    }
    \caption{\label{fig:g2_same_num}
        \textit{Off-diagonal number correlations when 1/3 of the atoms have escaped.}
        Same plot as Fig.~\ref{fig:g2_for_U}, but for a fixed number of escaped atoms rather than a fixed time: (a)-(f) $U=1,2,3,4,5,6$. Near the critical region, $U=3$, the positive correlations in the escape region begin to damp out, and are entirely lost in the Mott insulating regime of $U=5$ to $U=6$, while the fork-like structures persist showing negative correlations between the escape region and the trap. Note negative (blue) and positive (red) correlations have separate scales to highlight the role of positive correlations more clearly.
    }
\end{figure}

In Sec.~\ref{sec:beyond} we emphasized the difference between the dependence of number fluctuations, quantum fluctuation rate, and quantum entropy on the tunneling time vs. the number of atoms escaped.  Therefore, as a complement to Fig.~\ref{fig:g2_for_U}, in Fig.~\ref{fig:g2_same_num} we show the difference between correlations in the superfluid, critical, and Mott insulating regimes when approximately 1/3 of the atoms have escaped from the trap. The fork-like structure that emerges near the critical regime, $U=3$ in Fig.~\ref{fig:g2_same_num}, has weaker negative correlations, but is still persistent. However, periodic structure in the positive correlations at the onset of macroscopic quantum tunneling around the critical region, $U=3,4$ in Fig.~\ref{fig:g2_for_U}(c,d), is no longer present. The rate of escaping atoms is much slower, producing only a faint positive correlation region in Fig.~\ref{fig:g2_same_num}(c,d). 
In the Mott insulating regime, $U=5,6$ in Fig.~\ref{fig:g2_same_num}, the negative correlation fork-like structures dominate the dynamics, with diagonal lines of negative correlation indicating that at each reflection time of $t=10,20,30,40,50,\ldots$ in the trap another atom has a chance of being emitted at the barrier.  An average negative correlation pattern like this will be built up over many experiments, in each of which a single atom either is or is not emitted.

\subsection{Preliminary evidence of a pulsed and continuous-wave correlation atom laser}
\label{ssec:laser}

Finally, we want to emphasize an application of this work to atom lasers.  The orignal concept of the atom laser was continuous-wave and emphasized the emission of atoms through a hole in a harmonic trap created by a localized state transition in the atoms~\cite{bloch1999atom}.  This is a classical hole -- no quantum tunneling was involved.  It was subsequently shown that attractive interactions could be used to create a pulsed solitonic atom laser~\cite{carr2004pulsed,rodas2005controllable}.  Although the emphasis in these works was on number density, the correlations in the tail for an atom laser were measured experimentally in~\cite{ottl2005correlations}, leading to the field of atom interferometry and the remarkable observation of up to 10th order phase correlators~\cite{langen2015experimental}.  It was subsequently suggested that tunneling could cause fragmentation, or condensation into multiple modes, in attractive BECs in particular~\cite{lode2012interacting}.

\begin{figure}
    \subfloat{
        \includegraphics[width=0.97\linewidth]{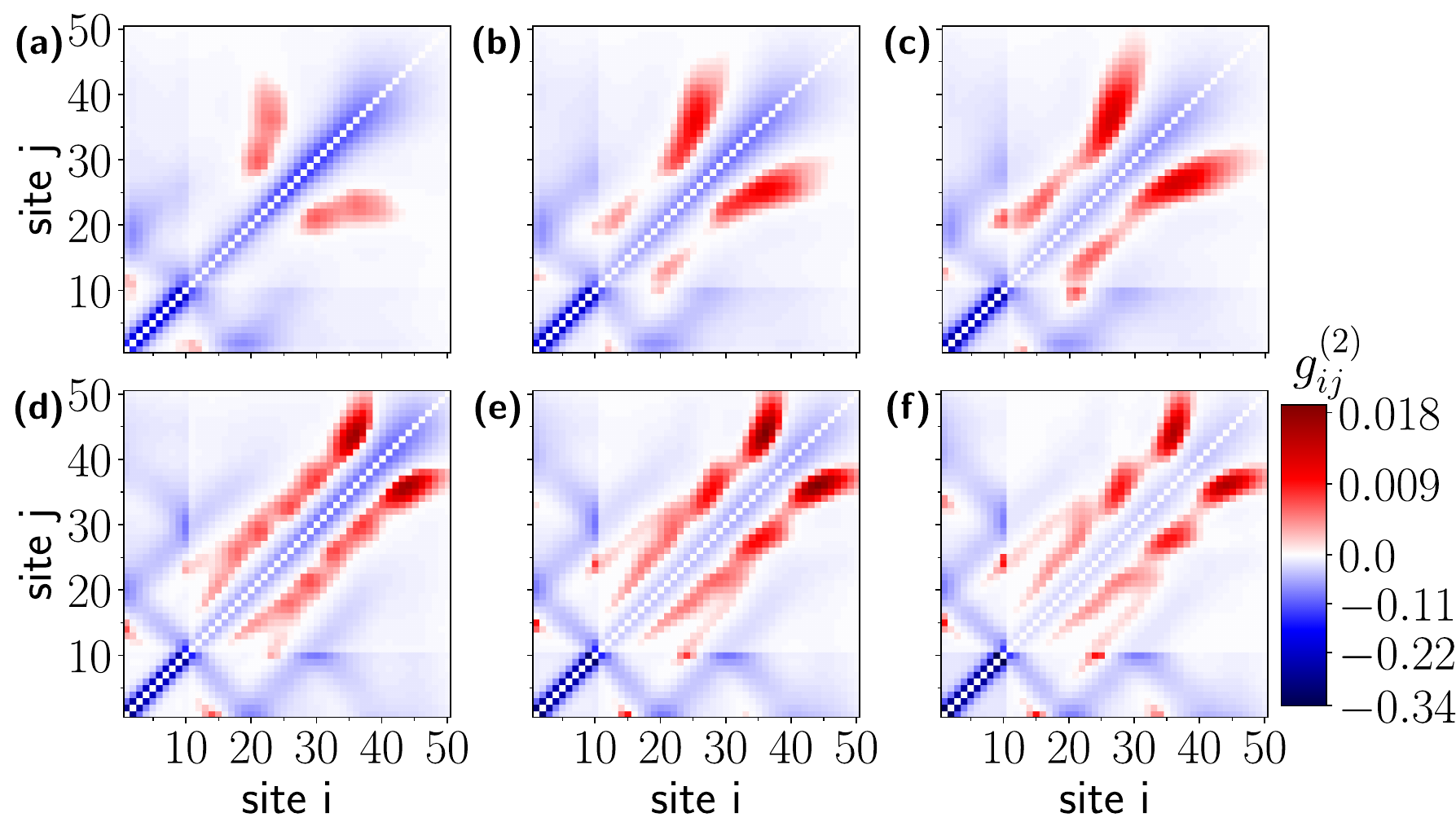}
    }
    \caption{\label{fig:g2_for_barrier}
        \textit{Preliminary demonstration of a correlation atom laser.}
        Barrier height and interactions are used to control pulsed and continuous wave correlation emissions. Interactions $U=1$ (top row) and $U=2$ (bottom row).  Barrier heights $h/U=1.0, 1.5, 2.0$ (columns left to right). All at time $t=20$. Larger barriers do not decrease the correlation strength, but instead produce visibly different positive correlation patterns in the escape region.
    }
\end{figure}

In Sec.~\ref{ssec:laser} we observed the creation of strong positive correlations in the escaped region in the superfluid region.  This effect can be further enhanced and focused by controlling the barrier height and interaction strength, creating a correlation atom laser. Such atom laser concepts have potential use in the field of atomtronics~\cite{seaman2007atomtronics}, where the flow of information may occur not only in currents and densities but also in higher order fluctuations.  In Fig.~\ref{fig:g2_for_barrier} we show how the barrier can be used to control off-diagonal number correlations.  For a low barrier and weak interactions the positive correlations appear in bursts, similar to the ``blips'' observed in the mean-field semiclassical limit of~\cite{dekel2007temporal} but here seen in a higher order off-diagonal quantity.  However, higher barriers and stronger interactions create extended and structured regions of positive correlation which flow continuously through the escape region.  Thus one can tune from pulsed to continuous-wave positive correlations by raising the barrier and tuning interaction within the superfluid regime.

We emphasize that the results of this section are preliminary and merit further detailed exploration of different pulsed and continuous-wave regimes, as this Article is not intended to focus on device applications.  For instance, our pulses range from about 2$\times$3 sites to about $3\times 8$ sites, indicating the barrier can used to shape as well as localize various off-diagonal correlation structures.  The axes of interactions vs. barrier size vs. initial filling factors all require a detailed study.

\section{Conclusions}
\label{sec:Conclusions}

We have demonstrated that two quantum phases, the superfluid and Mott insulator, show distinct macroscopic quantum tunneling escape dynamics.  These dynamics show many surprising features beyond previously known tunneling regimes, such as the here newly introduced quantum fluctuation rate to characterize quantum correlations which prove to be key to the tunneling process.  To demonstrate this new regime of macroscopic quantum tunneling, we evolved an entangled initial state of the Bose-Hubbard Hamiltonian modeling cold atoms in optical lattice quantum simulators trapped behind a narrow barrier of controlled height.  We found the effects persisted even in the mesoscopic regime of 10 particles on 10 sites accessible to many present or near-term quantum simulator platforms beyond cold atoms~\cite{altman2021quantum}, and are therefore experimentally realizable.

The subsequent dynamics were first characterized by a measure drawn from standard tunneling, namely the atom number escape rate.  In single-particle tunneling this is well-known to be a decaying exponential.  We found instead a highly non-exponential decay which depends strongly on the quantum phase.  In the weakly-interacting superfluid regime tunneling dynamics were found to be wave-like, with coherent interference patterns in the escape region, and decay was rapid and non-exponential, leading to a rapidly diminishing and non-constant tunneling rate. This is interpreted as due to the single-particle energy (equivalent to a chemical potential) dropping relative to the barrier height.  In the strongly-interacting Mott insulator regime we found particle-like tunneling, which was interpreted as suppression of two-particle tunneling events by the Mott gap.  The tunneling rate was found to also be non-constant, was an order of magnitude smaller despite stronger repulsive interactions, and decreased nearly linearly.  Overall, this effect is caused by the resistance of a Mott insulator to particle flow or mass current.  However, to understand these surprisingly non-exponential behaviors better it was necessary to go beyond single-particle measures and examine correlations.  This again emphasizes the role the quantum phase plays in the tunneling dynamics, as quantum phases are best characterized by quantum correlations.

We thus defined a new tunneling rate, the \emph{quantum fluctuation rate}, which can be used to characterize tunneling of fluctuations beyond the semiclassical picture.  We found that while the superfluid always has a positive rate, the Mott insulator at first has a negative rate, during which fluctuations actually increase, before they again decrease as the trap empties.  We explained this effect using the weakly and strongly interacting limits of the BHH, showing that when atoms tunnel one-by-one for strong interactions, fluctuations are maximized when about half the atoms have tunneled.  We then went on to examine quantum entropy created in the trapped atoms during the tunneling escape process, where we found that while in the superfluid regime entropy is maximized when about half the atoms have tunneled, whereas in the Mott insulator regime twice the amount of entropy is created and this occurs when only about one-quarter of the atoms have tunneled.  

Finally, beyond the quantum fluctuation rate we emphasized the quantum many-body nature of the tunneling process by studying off-diagonal second-order number correlations.  We showed the superfluid phase creates positive correlations in the escape region, similar to the bunching effect seen in other contexts, from entangled electrons~\cite{burkard2000noise} to the quark-gluon plasma~\cite{harris1996search}.  The Mott insulator phase leads to negative correlations between the trap and the escape region, demonstrating the fermionic character of tunneling, and further highlighting the role of the gap.  Only at very late times when the the quantum fluctuation rate again turns positive have enough particles been emitted that some of the superfluid character is recovered.

Applications of quantum phases in tunneling devices have been suggested in the context of Josephson junctions~\cite{mclain2018high,mclain2018quantum}.  The escape dynamics considered here present a generalization of the concept of an atom laser~\cite{bloch1999atom}, where correlations have already been measured explicitly~\cite{ottl2005correlations}.  Our work thus offers the possibility of future technological applications in guiding and controlling entangled quantum matter from atomtronics~\cite{seaman2007atomtronics} to quantum information science.  In particular, we offered a very preliminary demonstration of pulsed and continuous-wave correlation atom laser regimes controlled by barrier height and interaction strength, which bear further investigation.  Likewise, the Mott insulator quantum phase can used to control emission of atoms one-by-one rather than collectively.  

In future investigations, exploration of macroscopic tunneling dynamics of a variety of quantum phases presents itself as a natural growth of the work here.  From the Fermi-Hubbard model to Ising models to Heisenberg models to exotic XYZ magnetism and many other phases of quantum matter, quantum simulators~\cite{altman2021quantum} can directly access macroscopic quantum tunneling dynamics.  Especially when such calculations are inaccessible on a classical computer the many quantum simulator platforms offer an exciting opportunity to answer such fundamental questions in quantum dynamics as the nature of macroscopic quantum tunneling.  

\section*{Acknowledgments}

We acknowledge useful conversations with Joseph Glick, Matthias Weidemueller, and Xinxin Zhao.  This work was performed in part with support by the NSF
under grants OAC-1740130, CCF-1839232, PHY-1806372, and DGE-2125899; and in conjunction with
the QSUM program, which is supported by the Engineering and Physical Sciences
Research Council grant EP/P01058X/1.

%
%
%
%


\bibliography{Mott_refs}

\end{document}